\documentclass[smallextended]{svjour3}       % onecolumn (second format)
\smartqed  % flush right qed marks, e.g. at end of proof
\usepackage{amsmath,amssymb,amsfonts}
\usepackage{algorithmic}
\usepackage{graphicx}
\usepackage{textcomp}
\usepackage{xcolor}

\usepackage{balance}
\usepackage[procnumbered,ruled,vlined,linesnumbered]{algorithm2e}
\usepackage{array}
\usepackage{booktabs}
\usepackage{multirow}
\usepackage{threeparttable}
\usepackage{makecell}
\usepackage{siunitx}
\usepackage{stfloats}
\usepackage{tabularx}
\usepackage{subfigure}
\usepackage{enumerate}
\usepackage{hyperref}       % hyperlinks
\usepackage{url}            % simple URL typesetting
\def\BibTeX{{\rm B\kern-.05em{\sc i\kern-.025em b}\kern-.08em
    T\kern-.1667em\lower.7ex\hbox{E}\kern-.125emX}}
\usepackage{pdfpages}

\usepackage{natbib}
\setcitestyle{authoryear,aysep={},open={(},close={)}}
\DontPrintSemicolon
\SetKw{KwAnd}{and}
\SetFuncSty{textsc}
\SetKwInOut{Input}{Input\ \ \ \ }
\SetKwInOut{Output}{Output}

\def\calE{\mathcal{E}}
\def\calGtil{\mathcal{\tilde{G}}}
\def\calG{\mathcal{G}}
\def\calN{\mathcal{N}}
\def\calV{\mathcal{V}}

\newcommand\ee{\boldsymbol{\mathit{e}}}

\newcommand\pp{\boldsymbol{\mathit{p}}}

\renewcommand\ss{\boldsymbol{\mathit{s}}}
\def\tt{\boldsymbol{\mathit{t}}}

\newcommand\yy{\boldsymbol{\mathit{y}}}
\newcommand\zz{\boldsymbol{\mathit{z}}}
\newcommand\xx{\boldsymbol{\mathit{x}}}

\newcommand\xxtil{\boldsymbol{\mathit{\tilde{x}}}}
\newcommand\zztil{\boldsymbol{\mathit{\tilde{z}}}}

\renewcommand\AA{\boldsymbol{\mathit{A}}}

\newcommand\DD{\boldsymbol{\mathit{D}}}

\newcommand\II{\boldsymbol{\mathit{I}}}

\newcommand\MM{\boldsymbol{\mathit{M}}}
\newcommand\LL{\boldsymbol{\mathit{L}}}
\newcommand\PP{\boldsymbol{\mathit{P}}}

\newcommand\LLtil{\boldsymbol{\mathit{\tilde{L}}}}

\newcommand\PPtil{\boldsymbol{\mathit{\tilde{P}}}}
\newcommand\calLL{\boldsymbol{\mathit{\mathcal{L}}}}
\newcommand\calLLtil{\boldsymbol{\mathit{\mathcal{\tilde{L}}}}}
\newcommand\ALPHA{\boldsymbol{\Lambda}}

\newcommand{\one}{\mathbf{1}}
\newcommand{\zero}{\mathbf{0}}
\def\eps{\epsilon}

\newcommand\bbeta{\boldsymbol{\beta}}
\newcommand\eeps{\boldsymbol{\eps}}

\newcommand{\kh}[1]{\left(#1\right)}
\def\norm#1{\left\| #1 \right\|}
\def\smallnorm#1{\| #1 \|}

\begin{document}

% \author{
% \IEEEcompsocitemizethanks{\IEEEcompsocthanksitem Zuobai Zhang, Wanyue Xu and Zhongzhi Zhang are with the Shanghai Key Laboratory of Intelligent Information Processing, School of Computer Science, Fudan University, Shanghai 200433, China. Zhongzhi~Zhang is also with the Shanghai Engineering Research Institute of Blockchains,  Fudan University, Shanghai 200433, China. (e-mail: ;  ).\protect\\
% \IEEEcompsocthanksitem Guanrong Chen is with the Department of Electrical Engineering, City University of Hong Kong, Hong Kong SAR, China (e-mail:).}
% %\thanks{(Corresponding author: Zhongzhi Zhang)}
% }% <-this % stops a space

\title{Opinion Dynamics in Social Networks Incorporating Higher-Order Interactions %\thanks{Grants or other notes
}


\author{Zuobai Zhang \and 
        Wanyue Xu \and 
        Zhongzhi Zhang \and
        Guanrong Chen%etc.
}


\institute{Zuobai Zhang \and
           Wanyue Xu \and
           Zhongzhi Zhang \at
          Shanghai Key Laboratory of Intelligent Information
	Processing, School of Computer Science, Fudan University, Shanghai 200433, China \\
           \email{zhangzz@fudan.edu.cn}
           \and
           Guanrong Chen \at
           Department of Electronic Engineering, City University of Hong Kong, Hong Kong SAR, China
}

\date{Received: date / Accepted: date}

\maketitle

\begin{abstract}
	The issue of opinion sharing and formation has received considerable attention in the academic literature, and a few models have been proposed to study this problem. 
	However, existing models are limited to the interactions among nearest neighbors, with those second, third, and higher-order neighbors only considered indirectly, despite the fact that higher-order interactions occur frequently in real social networks. 
	In this paper, we develop a new model for opinion dynamics by incorporating long-range interactions based on higher-order random walks that can explicitly tune the degree of influence of higher-order neighbor interactions. We prove that the model converges to a fixed opinion vector, which may differ greatly from those models without higher-order interactions. Since direct computation of the equilibrium opinion is computationally expensive, which involves the operations of huge-scale matrix multiplication and inversion, we design a theoretically convergence-guaranteed estimation algorithm that approximates the equilibrium opinion vector nearly linearly in both space and time with respect to the number of edges in the graph. We conduct extensive experiments on various social networks, demonstrating that the new algorithm is both highly efficient and effective.
\keywords{Opinion dynamics \and social network \and computational social science \and random walk \and spectral graph theory}
\end{abstract}

\section{Introduction}

Recent years have witnessed an explosive growth in social media and online social networks, which have increasingly become an important part of our lives~\citep{SmCh08}. For example, online social networks can increase the diversity of opinions, ideas, and information available to individuals~\citep{Ki11,LeChKiKi14}. At the same time, people may use online social networks to broadcast information on their lives and their opinions about some topics or issues to a large audience. It has been reported that social networks and social media have resulted in a fundamental change of ways that people share and shape opinions~\citep{DaGoMu14,FoPaSk16,AuFeGr18}.  Recently, there have been a concerted effort to model opinion dynamics in social networks, in order to understand the effects of various factors on the formation dynamics of opinions~\citep{JiMiFrBu15,DoZhKoDiLi18,AnYe19}.

One of the popular opinion dynamics models is the Friedkin-Johnsen (FJ) model~\citep{FrJo90}. Although  simple and succinct, the FJ model can capture complex behavior of real social groups by incorporating French's ``theory of social power''~\citep{Fr56}, and thus has been extensively studied. A sufficient condition for the stability of this standard model was obtained in~\citep{RaFrTeIs15}, the average innate opinion was estimated in~\citep{DaGoPaSa13}, and the unique equilibrium expressed opinion vector was derived in~\citep{DaGoPaSa13,BiKlOr15}. Some explanations of this natural model were consequently explored from different perspectives~\citep{GhSr14,BiKlOr15}. In addition, based on the  FJ opinion dynamics model, some social phenomena have been quantified and studied~\cite{XuBaZh21}, including polarization~\citep{MaTeTs17,MuMuTs18},  disagreement~\citep{MuMuTs18}, conflict~\citep{ChLiDe18}, and controversy~\citep{ChLiDe18}.  Moreover, some optimization problems~\citep{AbKlPaTs18} for the   FJ model were also investigated, such as opinion  maximization~\citep{GiTeTs13}.

Other than studying the properties, interpretations and related quantities of the FJ model, many extensions or variants of this popular model have been developed~\citep{JiMiFrBu15}. In~\citep{AbKlPaTs18}, the impact of susceptibility to persuasion on opinion dynamics was analyzed by introducing a resistance parameter to modify the FJ model. In~\citep{SeGrSqRa19}, a varying peer-pressure coefficient was introduced to the FJ model, aiming to explore the role of increasing peer pressure on opinion formation. In~\citep{ChMu20}, the FJ model was augmented to include algorithmic filtering, to analyze the effect of filter bubbles on polarization. Some multidimensional extensions were developed for the FJ model~\citep{Fr15, PaPrTeFr15,PaPrTeFr17, FrPrTePa16}, extending the scalar opinion to vector-valued opinions corresponding to several settings, either independent~\citep{Fr15} or interdependent~\citep{PaPrTeFr15,PaPrTeFr17, FrPrTePa16}.

The above related works for opinion dynamic models provide deep insights into the understanding of opinion formulation, since they grasped various important aspects affecting opinion shaping, including individual's attributes, interactions among individuals, and opinion update mechanisms. However, existing models consider only the interactions among the nearest neighbors,
allowing interactions with higher-order neighbors only indirectly via their immediate neighbors,
%neglecting those interactions among second-order, third-order, and higher-order nearest neighbors, 
in spite of the fact that this situation is commonly encountered in real natural~\citep{ScLiRoLaStBe02} and  %SeEh95,
social~\citep{GhEbGa13,lyuYuanWang22} networks. In a real natural example~\citep{ScLiRoLaStBe02}, it is shown as the long-jump spanning multiple lattice spacings, which plays a dominating role in the diffusion of these molecules. In a social network example~\citep{GhEbGa13,lyuYuanWang22}, an individual can make use of the local, partial, or global knowledge corresponding to his direct, second-order, and even higher-order neighbors to search for opinions about a concerned issue or to diffuse information and opinions in an efficient way. It has been suggested by many existing theories and models that long ties are more likely to persist than other social ties, and that many of them constantly function as social bridges \citep{lyuYuanWang22}. \cite{ScHe22} defined a higher-order Deffuant model~\citep{DeNeAm00}, generalizing the original pairwise interaction model for bounded-confidence opinion-dynamics to interactions involving a group of agents.
To date, there is  still a lack a comprehensive higher-order FJ opinion dynamics model on social networks, although it has been observed that long-range non-nearest-neighbor interactions could play a fundamental role in opinion dynamics.

%\noindent
In this paper, we make a natural extension of the classical FJ opinion dynamics model to explicitly incorporate the higher-order interactions between individuals and their non-nearest neighbors  by leveraging higher-order random walks. We prove that the higher-order model converges to a unique equilibrium opinion vector, provided that each individual has a non-zero resistance parameter measuring his susceptibility to persuasion. We show that the equilibrium opinions of the higher-order FJ model differ greatly from those of the classical FJ model, demonstrating that higher-order interactions have a significant impact on  opinion dynamics.

Basically, the equilibrium opinions of the higher-order FJ model on a graph are the same as those of the standard FJ model on a corresponding dense graph with a loop at each node. That is, at each time step, every individual updates his opinion according to his innate opinion, as well as the currently expressed opinions of his nearest neighbors on the dense graph. Since the transition matrix of the dense graph is a combination of the powers of that on the original graph, direct construction of the transition matrix for the dense graph is computationally expensive. To reduce the computation cost, we construct a sparse matrix, which is spectrally close to the dense matrix, nearly linearly in both space and time with respect to the number of edges on the original graph. This sparsified matrix maintains the information of the dense graph, such that the difference between the equilibrium opinions on the dense graph and the sparsified graph is negligible.

Based on the obtained sparsifed matrix, we further introduce an iteration algorithm, which has a theoretical convergence and can approximate the equilibrium opinions of the higher-order FJ model quickly. Finally, we perform extensive experiments on different networks of various scales, and show that the new algorithm achieves high efficiency and effectiveness. Particularly, this algorithm is scalable, which can approximate the equilibrium opinions of the second-order FJ on large graphs with millions of nodes. It is expected that the new model  sheds light on  further understanding of opinion formation, and that the new algorithm can be helpful for  various applications, such as the computations of polarization and disagreement in opinion dynamics.

A preliminary version of our work has been published in~\citep{ZhXuZhCh20}. In this paper, we extend our preliminary results in several directions. First, we present proof details previously omitted in~\citep{ZhXuZhCh20} for several important theorems, including the convergence analysis and approximation error bound of the proposed algorithm. Second, we add an illustrative example in Section~\ref{HOFJ}, in order to better understand and demonstrate the difference between the traditional FJ model and the higher-order model. Finally, we provide additional experimental results for different innate opinion distributions and provide a thorough parameter analysis in Section~\ref{sec:exp}.

\section{Preliminaries}

In this section, some basic concepts in graph and matrix theories,  as well as the  Friedkin-Johnsen (FJ) opinion dynamics model are briefly reviewed.%,  some quantitelectrical networks and biharmonic distance.

\subsection{Graphs and Related  Matrices}

Consider a simple, connected, undirected social network (graph) $\calG=(\calV,\calE)$, where $\calV=\{1,2,...,n\}$ is the set of $n$ agents and $\calE=\{(i,j)|i,j\in\calV\}$ is the set of $m$ edges describing relations among nearest neighbors.
The topological and weighted properties of $\calG$ are encoded in its adjacency matrix $\AA=(a_{ij})_{n\times n}$, where $a_{ij}=a_{ji}=w_e$ if $i$ and $j$ are linked by an edge $e=(i,j)\in\calE$ with weight $w_e$, and $a_{ij}=0$ otherwise.
Let $\calN_i=\{j|(i,j)\in\calE\}$ denote the set of neighbors of node $i$ and $d_i=\sum_{j\in\calN_i} a_{ij}$ denote the degree of $i$.
The diagonal degree matrix of graph $\calG$ is defined to be $\DD = {\rm diag}(d_1,d_2,...,d_n)$, and the Laplacian matrix of $\calG$ is $\LL=\DD-\AA$.
Let $\ee_i$ denote the $i$-th standard basis vector of appropriate dimension.
Let $\one$ ($\zero$) be the vector with all entries being ones (zeros).
Then, it can be verified that $\LL\one=\zero$.
The random walk transition matrix for $\calG$ is defined as $\PP=\DD^{-1}\AA$, which is row-stochastic (i.e., each row-sum equals $1$). % i.e., each entry of $\PP$ is non-negative and the sum of all entries in each row is $1$.

\subsection{Norms of a Vector or Matrix}

For a non-negative vector $\xx$, $x_{\max}$ and $x_{\min}$ denote the maximum and minimum entry, respectively.
For an $n\times n$ matrix $\AA$, $\sigma_i(\AA), i=1,2,...,n$ denote its singular values.
Given a vector $\xx$, its $2$-norm is defined as $\smallnorm{\xx}_2=\sqrt[2]{\sum_i{|x_i|^2}}$ and the $\infty$-norm is defined as $\smallnorm{\xx}_{\infty}=\max_i{|x_i|}$.
It is easy to verify that $\smallnorm{\xx}_{\infty}\le\smallnorm{\xx}_2\le\sqrt{n}\smallnorm{\xx}_{\infty}$ for any $n$-dimensional vector $\xx$.
For a matrix $\AA$, its $2$-norm is defined to be $\smallnorm{\AA}_2=\max_{\xx} \smallnorm{\AA\xx}_2/\smallnorm{\xx}_2$.
By definition, $\smallnorm{\AA\xx}_{2}\le\smallnorm{\AA}_2\smallnorm{\xx}_2$.
It is known that the $2$-norm of the matrix $\AA$ is equal to its maximum singular value $\sigma_{\max}$, satisfying $\smallnorm{\xx^\top\AA\yy}_2\le\sigma_{\max}\smallnorm{\xx}_2\smallnorm{\yy}_2$ for any vectors $\xx$ and $\yy$~\citep{GoVa12}.

\subsection{Friedkin-Johnsen Opinion Dynamics Model}

The Friedkin-Johnsen (FJ) model is a classic opinion dynamics model~\citep{FrJo90}. For a specific topic, the FJ model assumes that each agent $i\in\calV$ is associated with an {\em innate opinion} $s_i\in[0,1]$, where higher values signify more favorable opinions, and a {\em resistance parameter} $\alpha_i\in(0,1]$ quantifying the agent's {\em stubbornness}, with a higher value corresponding to a lower tendency to conform with his neighbors' opinions. Let $\xx^{(t)}$ denote the opinion vector of all agents at time $t$, with element $x_i^{(t)}$ representing the opinion of agent $i$ at that time. At every timestep, each agent updates his opinion by taking a convex combination of his innate opinion and the average of the expressed opinion of his neighbors in the previous timestep. Mathematically, the opinion of agent $i$ evolves according to the following rule:
\begin{equation}
    x_i^{(t+1)}=\alpha_i s_i+(1-\alpha_i)\frac{\sum_{j\in\calN_i}a_{ij}\cdot x_j^{(t)}}{d_i}.
\end{equation}
The evolution rule can be rewritten in matrix form as
\begin{equation}
\label{eq:update}
    \xx^{(t+1)}=\ALPHA \ss+(\II-\ALPHA)\PP\xx^{(t)},
\end{equation}
where $\ALPHA$  denotes the diagonal matrix ${\rm diag}(\alpha_1,\alpha_2,...,\alpha_n)$, and $\II$  is the identity matrix.

It has been proved~\citep{DaGoPaSa13} that the above opinion formation process converges to a unique equilibrium $\zz$ when $\alpha_i>0$ for all $i\in\calV$. The equilibrium vector $\zz$ can be obtained as the unique fixed point of equation~\eqref{eq:update}, i.e.,
\begin{equation}
    \zz=\kh{\II-\kh{\II-\ALPHA}\PP}^{-1}\ALPHA\ss\,.
\end{equation}
The $i$th entry $z_i$ of $\zz$ is the {\em expressed opinion} of agent $i$.

A straightforward way to calculate the equilibrium vector $\zz$ requires inverting a matrix, which is expensive and intractable for large networks. In~\citep{ChLiSo19}, the iteration process of the opinion dynamics model is used to obtain an approximation of vector $\zz$, which has a theoretical guarantee of convergence. The method is very efficient, scalable to networks with millions of nodes.

\section{Higher-order Opinion Dynamics Model}

The classical FJ model has many advantages; for example, it captures some complex human behavior in social networks. However, this model considers only the interactions among nearest neighbors, without explicitly considering the higher-order interactions existing in social networks and social media. To fix this deficiency, in this section, we generalize the FJ model to a higher-order setting by using the random walk matrix polynomials describing higher-order random walks.

\subsection{Random Walk Matrix Polynomial}

For a network $\calG$, its random walk matrix polynomial is defined as follows~\citep{ChChLiPeTe15b}:
\begin{definition}
Let $\AA$ and $\DD$ be, respectively, the adjacency matrix and diagonal degree matrix of a graph $\calG$. For a non-negative vector $\bbeta=(\beta_1,\beta_2,...,\beta_T)$ satisfying $\sum_{r=1}^T \beta_r=1$, the matrix
\begin{equation}
\label{eq:LLbeta}
    \LL_{\bbeta}(\calG)=\DD-\sum_{r=1}^T \beta_r\DD\kh{\DD^{-1}\AA}^r
\end{equation}
is a $T$-degree random walk matrix polynomial of $\calG$.
\end{definition}

The Laplacian matrix $\LL$ is a particular case of $\LL_{\bbeta}(\calG)$, which can be obtained from $\LL_{\bbeta}(\calG)$ by setting $T=1$ and $\beta_1=1$.
In fact, it can be proved that, for any $\bbeta$, there always exists a graph $\calG'$ with loops, whose Laplacian matrix is $\LL_{\bbeta}(\calG)$, as characterized by the following theorem.
\begin{theorem}[Proposition 25 in~\citet{ChChLiPeTe15b}]
The random walk matrix polynomial $\LL_{\bbeta}(\calG)$ is a Laplacian matrix.
\end{theorem}

Define matrix $\LL_{\calG_r}=\DD-\DD\kh{\DD^{-1}\AA}^r$, which is a particular case of matrix $\LL_{\bbeta}(\calG)$  corresponding to $T=r$ and $\beta_r=1$. In fact, $\LL_{\calG_r}$ is the Laplacian matrix of graph $\calG_r$, constructed from graph $\calG$ by performing $r$-step random walks on graph $\calG$. The $ij$-th element of the adjacency matrix $\AA_{\calG_r}$ for graph $\calG_r$ is equal to the product of the degree $d_i$ for node $i$ in $\calG$ and the probability that a walker starts from node $i$ and ends at node $j$ after performing $r$-step random walks in $\calG$. Thus, the matrix polynomial $\LL_{\bbeta}(\calG)$ is a combination of matrices $\LL_{\calG_r}$ for $r=1,2,...,T$.

Based on the  random walk matrix polynomials, one can define a generalized transition matrix $\PP^\ast=\PP^\ast_{\bbeta}$  for  graph $\calG$ as follows.
\begin{definition}
Given an undirected weighted graph $\calG$ and a coefficient vector $\bbeta=(\beta_1,\beta_2,...,\beta_T)$ with $\sum_{r=1}^T\beta_r=1$, the matrix
\begin{equation}
    \PP^\ast_{\bbeta}=\sum_{r=1}^T\beta_r\PP^r
    =\II-\DD^{-1}\LL_{\bbeta}(\calG)
\end{equation}
is a $T$-order transition matrix of $\calG$ with respect to vector $\bbeta$.
\end{definition}
Note that the generalized transition matrix $\PP^\ast$ for  graph $\calG$  is actually the transition matrix for another graph $\calG'$.

\subsection{Higher-Order FJ Model}\label{HOFJ}

To introduce the higher-order FJ model, first modify the update rule in equation~\eqref{eq:update} by replacing $\PP$ with $\PP^\ast$. In other words, the opinion vector evolves as follows:
\begin{align}
    \label{eq:upma}
    \xx^{(t+1)}&=\ALPHA \ss+(\II-\ALPHA)\PP^\ast\xx^{(t)}\\
    &=\ALPHA \ss+(\II-\ALPHA)\left[\beta_1\PP+\beta_2\PP^2+...+\beta_T\PP^T\right]\xx^{(t)}.\nonumber
\end{align}
In this way, individuals update their opinions by incorpating those of their higher-order neighborhoods  at each timestep.
Moreover, by adjusting the coefficient vector $\bbeta$, one can choose different weights for neighbors of different orders.

Note that for the case of $\PP^\ast=\PP$, the higher-order FJ model is reduced to the classic FJ model.
While for the case of $\PP^\ast \neq \PP$, the higher-order FJ model can lead to very different results, in comparison with the standard FJ model, as shown in the following example.

%%%%%%%%%%%%%%%%%%%%%%%%%%%%%%%%%%%%%%%%%%%%%%%%
\begin{figure}[h]
	\centering
	\includegraphics[width=0.25\linewidth]{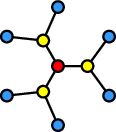}
	\caption{A tree with ten nodes.}
	\label{fig:Cayley}
\end{figure}
%%%%%%%%%%%%%%%%%%%%%%%%%%%%%%%%%%%%%%%%%%%%%%%%

\noindent\textbf{Example.}
Consider the tree shown in Figure~\ref{fig:Cayley}. %, which is constructed by first generating a star graph with four nodes and then creating two adjunct nodes for each periphery node.
The nodes from the center to the periphery are colored in red, yellow and blue, respectively. Suppose that for the red node its $(s_i,\alpha_i)$ are given by $(1,1)$, implying that the center node has a favorable opinion and is insusceptible to be persuaded by others. And, suppose that the yellow and blue nodes have values $(0,0.6)$ and $(0,0.01)$, respectively. Then, calculate the equilibrium opinion vector for the following three cases:

\begin{enumerate}[I.]
	\item $\beta_1=1,\beta_2=0$. This case corresponds to the classic FJ model. At every timestep, the opinion of each node is influenced by the opinions of its nearest neighbors. At equilibrium, the expressed opinions of red, yellow and blue nodes are $1$, $0.181$ and $0.179$, respectively. This is consistent with the intuition, since the red and yellow nodes are stubborn and thus prone to their innate opinions, while the blue nodes are susceptible to their neighboring nodes, the yellow ones.
	\item $\beta_1=0,\beta_2=1$. This case is associated with the second-order FJ model. In this case, only the influences of the second-order neighbors are considered. The equilibrium expressed opinions of the red, yellow and blue nodes are $1$, $0$ and $0.971$, respectively. This can be explained as follows. Since any yellow node is the second-order neighbor of the other two yellow nodes, they are influenced by each other, so they all stick to their innate opinions. In contrast, the blue nodes are highly affected by the center node.
	\item $\beta_1=\beta_2=0.5$. This is in fact a hybrid case of the above two cases, with interactions between a node and both of its first- and second-order neighbors with equal weights. For this case, the equilibrium opinions for red, yellow and blue nodes are $1$, $0.142$ and $0.351$, respectively. The opinion of each node lies between the opinions of the above two cases.
\end{enumerate}
In addition to the expressed opinion of an individual, for the above-considered three cases, the sum of their expressed opinions are also significantly different, which are equal to $2.617$, $6.826$ and $3.532$, respectively.

This example demonstrates that the interactions between nodes and their higher-order neighbors can have substantial impact on network opinions.
%The high-order FJ model gives us a way of modeling this impact.
Moreover, as will be seen in Section~\ref{sec:comp},  the higher-order interactions also strongly affect the opinion dynamics on real-world social networks.

\section{Convergence Analysis}

In this section, the convergence of the higher-order FJ model is analyzed.
It will be shown that if all $\alpha_i$ are positive, the model has a unique equilibrium and will converge to that equilibrium after sufficiently many iterations.
The high-level ideas of our proof are adapted from~\cite{DaGoPaSa13}.
%{\color{red} It should be noted that unlike traditional FJ model, the diagonal entries of the generalized transition matrix $\PP^\ast$ may be non-zero in our high-order FJ model, which leads to a similar but different proof.}

First, recall the Gershgorin Circle Theorem.

\begin{lemma}[Gershgorin Circle Theorem~\citet{Be65}]
\label{lem:GCT}
Given a square matrix $\AA\in\mathbb{R}^{n\times n}$, let $R_i=\sum_{j\neq i}|a_{ij}|$ be the sum of the absolute values of the non-diagonal entries in the $i$-th row and $D(a_{ii},R_i)\in\mathbb{C}$ be a closed disc centered at $a_{ii}$ with radius $R_i$.
Then, every eigenvalue of $\AA$ lies in at least one of the discs $D(a_{ii},R_i)$.
\end{lemma}

Now, the following main result is established.

\begin{theorem}
The higher-order FJ model defined in~\eqref{eq:upma} has a unique equilibrium if $\alpha_i>0$ for all $i \in\calV$.
\end{theorem}
\begin{proof}
Any equilibrium $\zz^\ast\in\mathbb{R}^n$ of~\eqref{eq:upma} must be a solution of the following linear system:
\begin{equation}
\label{eq:eq}
    \kh{\II-(\II-\ALPHA)\PP^\ast}\zz^\ast=\ALPHA\ss.
\end{equation}
Let $\MM=\II-(\II-\ALPHA)\PP^\ast$. It suffices to show that $\MM$ is non-singular.
First, it is obvious that every diagonal entry of $\MM$ is non-negative and every non-diagonal entry is non-positive, since every entry of $\PP^\ast$ lies in the interval $[0,1]$.
Thus, for any row $i$ of $\MM$, the sum of absolute values of its non-diagonal elements is
\begin{equation*}
    R_i=\sum_{j\neq i} |M_{ij}|
    =-\ee_i^\top\MM\one + M_{ii}
    =M_{ii}-\alpha_i>0,
\end{equation*}
where  $M_{ij}$  denotes the $(i,j)$th element of $\MM$.
Then, according to Lemma~\ref{lem:GCT}, every eigenvalue $\lambda$ of $\MM$ lies within the discs $\{z:|z-M_{ii}|\le M_{ii}-\alpha_i\}$.
Since $\alpha_i>0$, the set of all eigenvalues for $\MM$ excludes $0$.
Therefore, matrix $\MM$ is invertible, and thus the equilibrium is unique.
\end{proof}

Hence, $\zz^\ast = \kh{\II-(\II-\ALPHA)\PP^\ast}^{-1}\ALPHA\ss$ is the unique equilibrium of the opinion dynamics model defined by~\eqref{eq:upma}.
Next, it will be proved that after sufficiently many iterations, the new higher-order FJ model will converge to this equilibrium.

\begin{theorem} \label{theorem:Zstar}
If $\alpha_i>0$ for all $i\in\calV$, then the higher-order FJ model converges to its unique equilibrium $\zz^\ast = \kh{\II-(\II-\ALPHA)\PP^\ast}^{-1}\ALPHA\ss$.
\end{theorem}
\begin{proof}
Define the error vector $\eeps^{(t)}$ at the $t$-th iteration  as $\eeps^{(t)}=\xx^{(t)}-\zz^\ast$. 
It can be proved that as $t$ approaches infinity, the error term $\eeps^{(t)}$ tends to zero. Substituting~\eqref{eq:upma} into the formula of $\eeps^{(t)}$, one obtains $\eeps^{(t+1)}=(\II-\ALPHA)\PP^\ast\eeps^{(t)}$. In addition, because the sum of all the entries in each row of $\PP^\ast$ is one and $(1-\alpha_i)\in [0,1]$, the elements of $\eeps^{(t)}$ become smaller as $t$ grows. This can be explained as follows. Let $\eps^{(t+1)}_{\max}$ be the element of vector $\eeps^{(t+1)}$ that has the largest absolute value. Then,
\begin{align*}
    \left|\eps^{(t+1)}_{\max}\right| &= \max_{i=1,2,...,n}\left\{(1-\alpha_i)\left|\sum_{j=1}^n P^\ast_{ij}\eps^{(t)}_{j}\right|\right\}\\
    &\le \max_{i=1,2,...,n}\left\{(1-\alpha_i)\sum_{j=1}^n P^\ast_{ij}\left|\eps^{(t)}_{\max}\right|\right\}\\
    &= \max_{i=1,2,...,n}\left\{(1-\alpha_i)\left|\eps^{(t)}_{\max}\right|\right\}
    < \left|\eps^{(t)}_{\max}\right|,
\end{align*}
which completes the proof of the theorem.
\end{proof}

\section{Fast Estimation of equilibrium opinion vector}

To compute the equilibrium expressed opinion vector $\zz^*$ requires calculating matrix $\PP^\ast$ and inverting a matrix, both of which are time consuming. In general, for a network $\calG$, sparse or dense, its $r$-step random walk graph $\calG_r$ could be very dense. Particularly, for a small-world network with a moderately large $r$, its $r$-step random walk graph $\calG_r$ is a weighted and almost complete graph. This makes it infeasible to compute the generalized transition matrix $\PP^\ast$ for huge networks.

In this section, the spectral graph sparsification technique is utilized to obtain an approximation of matrix $\PP^\ast$. Then, a fast convergent algorithm is developed to approximate the expressed opinion vector $\zz^*$, which  avoids  matrix inverse operation. The pseudocode of this new algorithm is shown in Algorithm~\ref{alg:sparsify}.

\subsection{Random-Walk Matrix Polynomial Sparsification}
\label{sec:sgs}

First, we introduce the concept of spectral similarity and the technique of random-walk matrix polynomial sparsification.

\begin{definition}[Spectral Similarity of Graphs~\citet{SpSr11}]
Consider two weighted undirected networks $\calG=(\calV,\calE)$ and $\calGtil=(\calV,\tilde{\calE})$. Let $\LL$ and $\LLtil$ denote, respectively, their Laplacian matrices. Graphs $\calG$ and $\calGtil$ are $(1+\eps)$-spectrally similar if
\begin{equation}
(1-\eps)\cdot\xx^\top\LLtil\xx\le\xx^\top\LL\xx\le(1+\eps)\cdot\xx^\top\LLtil\xx,\quad \forall\xx\in\mathbb{R}^n.
\end{equation}
\end{definition}

Next, recall the sparsification algorithm~\cite{ChChLiPeTe15b}.
For a given graph $\calG=(\calV,\calE)$, start from an empty graph $\calGtil$ with the same node set $\calV$ and an empty edge set.
Then add $M$ edges into the sparsifier $\calGtil$ iteratively by a sampling technique.
At each iteration, randomly pick an edge $e=(u,v)$ from $\calE$  as an intermediate edge and an integer $r$ from $\{1,2,...,T\}$  as the length of the random-walk path.
To this end, run the \textsc{PathSampling}$(e,r)$ algorithm~\cite{ChChLiPeTe15b} to sample an edge by performing $r$-step random walks, and add the sample edge, together with its corresponding weight, into the sparsifier $\calGtil$. Note that multiple edges will be merged into a single edge by summing up their weights together. Finally, the algorithm generates a sparsifier $\calGtil$ for the original graph $\calG$ with no more than $M$ edges.

In~\cite{ChChLiPeTe15b}, an algorithm is designed to obtain a sparsifier $\calGtil$ with $O(n\eps^{-2}\log n)$ edges for $\LL_{\bbeta}(\calG)$, which consists of two steps:
The first step uses random walk path sampling to get an initial sparsifier with $O(Tm\eps^{-2}\log n)$ edges. The second step utilizes the standard spectral sparsification algorithm proposed in~\cite{SpSr11} to further reduce the edge number to $O(n\eps^{-2}\log n)$.  Since a sparsifier with $O(Tm\eps^{-2}\log n)$ edges is sparse enough for the present purposes, only the first step will be taken, while skipping the second step, to avoid unnecessary computations.

%%%%%%%%%%%%%%%%%%%%%%%%%%%%%%%%%%%%%%%%%%%%%
\begin{algorithm}
	\caption{\textsc{HODynamic}$(\calG,M,\ss,\bbeta,t)$}
	\label{alg:sparsify}
	\Input{
		$\calG$: a connected undirected graph; \\
		$M$: the number of edges in sparsifier; \\
		$\ss$: the innate opinion vector; \\
		$\bbeta$: the coefficient vector of random walk matrix polynomial;\\
		$t$: the number of iterations;\\
	}
	\Output{
		$\xxtil^{(t)}$: the approximate equilibrium vector;
		%$\calGtil$: a sparse graph with no more than $M$ edges, which is spectrally similar with $\calG$
	}
	$\calGtil=(\calV,\emptyset)$ \;
	\For{$i=1$ to $M$}{
		Randomly pick an edge $e=(u,v)\in\calE$  \;
		Select an integer $r$ from $\{1,2,...,T\}$ at uniform as the length of the random-walk path \;
		Randomly pick an integer $k\in\{1,2,...,r\}$\;
		Perform $(k-1)$-step random walk from $u$ to $u_0$\;
		Perform $(r-k)$-step random walk from $v$ to $u_r$\;
		Calculate $Z(\pp)$ along the length-$r$ path $\pp$ between node $u_0$ and node $u_r$ according to~\eqref{eq:Zp}\;
		%$u',v',Z\leftarrow$\textsc{PathSampling}$(e,r)$ \\
		Add an edge $(u_0,u_r)$ of weight $\frac{2rm\beta_r}{MZ}$ to $\calGtil$
	}
	$\PPtil=\II-\DD^{-1}\LLtil(\calGtil)$\;
	$\xxtil^{(0)}=\ss$\;
	\For{$i=1$ to $t$}{
		$\xxtil^{(i)} = \ALPHA\ss+(\II-\ALPHA)\PPtil\xxtil^{(i-1)}$\;
	}
	\Return $\xxtil^{(t)}$
\end{algorithm}
%%%%%%%%%%%%%%%%%%%%%%%%%%%%%%%%%%%%%%%%%%%%%

To sample an edge by performing $r$-step random walks, the procedure of \textsc{PathSampling} algorithm~\cite{ChChLiPeTe15b} is characterized in Lines 5-9 of Algorithm~\ref{alg:sparsify}. To sample an edge, first draw a random integer $k$ from $\{1,2,...,r\}$ and then perform, respectively, $(k-1)$-step and $(r-k)$-step walks starting from two end nodes of the edge $e=(u,v)$. This process samples a length-$r$ path $\pp=(u_0,u_1,...,u_r)$.
At the same time, compute
\begin{equation}
\label{eq:Zp}
    Z(\pp)=\sum_{i=1}^r \frac{2}{a_{u_{i-1},u_i}}.
\end{equation}
The algorithm returns the two endpoints of path $\pp$ as the sample edge $(u_0,u_r)$ and the quantity $Z(\pp)$ for the  calculation of weight.

\begin{theorem}[Spectral Sparsifiers of Random-Walk Matrix Polynomials~\cite{ChChLiPeTe15b}]
For a graph $\calG$ with random-walk matrix polynomial
\begin{equation}
    \LL_{\bbeta}(\calG) = \DD - \sum_{r=1}^T \beta_r \DD \kh{\DD^{-1}\AA}^r,
\end{equation}
where $\sum_{r=1}^T \beta_r=1$ and $\beta_r$ are non-negative, one can construct, in time $O(T^2m\eps^{-2}\log^2 n)$, a $(1+\eps)$-spectral sparsifier, $\LLtil$, with $O(n\eps^{-2}\log n)$ non-zeros.
\end{theorem}

Now one can approximate the generalized transition matrix using the Laplacian $\LLtil(\calGtil)$ of the sparse graph $\calGtil$:
\begin{equation}
    \PP^\ast = \II-\DD^{-1}\LL_{\bbeta}(\calG)
    \approx \II-\DD^{-1}\LLtil(\calGtil)
    =\PPtil^\ast.
\end{equation}

\paragraph{Complexity Analysis}
Regarding the time and space complexity of sparsification process of Algorithm~\ref{alg:sparsify}, the main time cost of sparsification (Lines 2-9) is the $M$ calls of the \textsc{PathSampling} routine. In \textsc{PathSampling}, it requires $O(\log n)$ time to sample a neighbor from the weighted network, and thus takes $O(r\log n)$ time to sample a length-$r$ path. Totally, the time complexity of Algorithm~\ref{alg:sparsify} is $O(MT\log n)$. As for space complexity, it takes $O(n+m)$ space to store the original graph $\calG$ and additional $O(M)$ space to store the sparisifier $\calGtil$. Thus, for  appropriate size $M$, the sparsifier is computable.

\subsection{Approximating the Equilibrium Opinion Vector via the Iteration Method}
\label{sec:iter}

With the spectral graph sparsification technique, it is possible to approximate $\PP^\ast$ with a sparse matrix. Nevertheless, directly computing the equilibrium still involves the matrix inverse operation, which is computationally expensive for large networks, such as those with millions of nodes. To approximate the equilibrium vector $\zz^\ast$ using the recurrence defined in~\eqref{eq:upma} and multiple iterations, in this section, we develop a convergent approximation algorithm. For this purpose, an important lemma is first introduced.
\begin{lemma}[Lemma 4 in~\cite{QiDoMaLiWaWaTa19}]
Let $\calLL=\DD^{-1/2}\LL\DD^{-1/2}$,  $\calLLtil=\DD^{-1/2}\LLtil\DD^{-1/2}$ and $\eps<0.5$.
Then all the singular values of $\calLLtil-\calLL$ satisfy that for all $i\in\{1,2,...,n\}$, $\sigma_i(\calLLtil-\calLL)<4\eps$.
\end{lemma}

% Since $\DD^{-1}\LL$ and $\DD^{-1}\LLtil$ are similar to $\calLL$ and $\calLLtil$, respectively,  the sets of eigenvalues for $\DD^{-1}\LL$ and $\calLL$ are identical. The same is true for $\DD^{-1}\LLtil$ and $\calLLtil$. Thus, the following result is obvious.
% \begin{corollary}
% \label{cor:sig}
% All the singular values of $\DD^{-1}\LLtil-\DD^{-1}\LL$ are smaller than $4\eps$, i.e., $\forall i\in\{1,2,...,n\}$, $\sigma_i(\DD^{-1}\LLtil-\DD^{-1}\LL)<4\eps$.
% \end{corollary}

Now, we are  in position to introduce a new iteration method for approximating the equilibrium vector $\zz^\ast$. First, set $\xxtil^{(0)}=\xx^{(0)}=\ss$. Then, in every timestep, update the opinion vector with the approximate transition matrix $\PPtil^\ast$, i.e., $\xxtil^{(t+1)}=\ALPHA \ss+(\II-\ALPHA)\PPtil^\ast\xxtil^{(t)}$. Let $\alpha_{\min}$ be the smallest value among all  $i\in\calV$. 
\begin{lemma}
\label{lem:xtil}
For every $t\ge 0$,
\begin{equation}
\label{eq:upb}
    \norm{\xxtil^{(t)}-\xx^{(t)}}_{\infty}
    \le \frac{4\eps\sqrt{nd_{\max}d^{-1}_{\min}}\cdot\kh{1-\alpha_{\min}}\left[1-\kh{1-\alpha_{\min}}^t\right]}{\alpha_{\min}}.
\end{equation}
\end{lemma}
\begin{proof}
Inequality~\eqref{eq:upb} is proved by induction.
The case of $j=0$ is trivial, since $\xxtil^{(0)}=\xx^{(0)}=\ss$.
Assume that~\eqref{eq:upb} holds for some integer $t$.
Then, it needs to show that~\eqref{eq:upb} also holds for $t+1$.
To this end, split $\smallnorm{\xxtil^{(t+1)}-\xx^{(t+1)}}_{\infty}$ into two terms by using triangle inequality:
\begin{align}
  &\quad  \norm{\xxtil^{(t+1)}-\xx^{(t+1)}}_{\infty}\\
    &\le\norm{\kh{\II-\ALPHA}\DD^{-1}\kh{\LL_{\bbeta}-\LLtil_{\bbeta}}\xx^{(t)}}_{\infty}\nonumber\\
    &+\norm{\kh{\II-\ALPHA}\kh{\II-\DD^{-1}\LLtil_{\bbeta}}\kh{\xxtil^{(t)}-\xx^{(t)}}}_{\infty}.
    \label{eq:sp}
\end{align}
For every coordinate of the first term in~\eqref{eq:sp}, an upper bound can be derived as follows:
\begin{align}
    &\left|\ee_i^\top\kh{\II-\ALPHA}\DD^{-1}\kh{\LL_{\bbeta}-\LLtil_{\bbeta}}\xx^{(t)}\right| \nonumber\\
    \le&(1-\alpha_{\min})\cdot\left|\ee_i^\top\DD^{-1}\kh{\LL_{\bbeta}-\LLtil_{\bbeta}}\xx^{(t)}\right| \nonumber\\
    \le&(1-\alpha_{\min})\cdot\sigma_{\max}\kh{\DD^{-1}\kh{\LL_{\bbeta}-\LLtil_{\bbeta}}}\left\|\ee_i\right\|_2\left\|\xx^{(t)}\right\|_2 \nonumber\\
    \le& (1-\alpha_{\min})\cdot \sigma_{\max}(\DD^{-1/2})\cdot \sigma_{\max}(\DD^{1/2})\cdot \sigma_{\max}\kh{\calLL_{\bbeta}-\calLLtil_{\bbeta}}\left\|\ee_i\right\|_2\left\|\xx^{(t)}\right\|_2\nonumber\\
    \le& \frac{4\eps(1-\alpha_{\min})d^{1/2}_{\max}}{d^{1/2}_{\min}}\sqrt{n},
    \label{eq:up1}
\end{align}
where the second inequality is obtained by using the inequality that $\xx^\top \AA \yy \le \sigma_{\max}(\AA) \|\xx\|_2 \|\yy\|_2$ for any matrix $\AA$,the third one follows $\|\DD^{-1}\kh{\LL_{\bbeta}-\LLtil_{\bbeta}}\|_2\le \|\DD^{-1/2}\|_2\|\DD^{1/2}\|_2\|\calLL_{\bbeta}-\calLLtil_{\bbeta}\|_2$ and the last inequality follows from  the fact that $x^{(t)}_i\le 1$  for all $i\in\{1,2,...,n\}$.

Next, consider the second term in~\eqref{eq:sp}.
One has
\begin{align}
    &\norm{\kh{\II-\ALPHA}\kh{\II-\DD^{-1}\LLtil_{\bbeta}}\kh{\xxtil^{(t)}-\xx^{(t)}}}_{\infty} \nonumber\\
    \le&\norm{\II-\ALPHA}_{\infty}\norm{\II-\DD^{-1}\LLtil_{\bbeta}}_{\infty}\norm{\xxtil^{(t)}-\xx^{(t)}}_{\infty}\nonumber\\
    =&\kh{1-\alpha_{\min}}\cdot\norm{\xxtil^{(t)}-\xx^{(t)}}_{\infty},
    \label{eq:up2}
\end{align}
where the equality is due to the fact that $\smallnorm{\II-\DD^{-1}\LLtil_{\bbeta}}_{\infty}=1$, which can be understood as follows. Since every entry of $\II-\DD^{-1}\LLtil_{\bbeta}$ is non-negative and $\LLtil_{\bbeta}\one=\zero$, one has
\begin{equation*}
    \norm{\II-\DD^{-1}\LLtil_{\bbeta}}_{\infty}
    =\norm{\kh{\II-\DD^{-1}\LLtil_{\bbeta}}\one}_{\infty}
    =1.
\end{equation*}

Substituting~\eqref{eq:up1} and~\eqref{eq:up2} into~\eqref{eq:sp}, one obtains
\begin{align*}
   &\quad \norm{\xxtil^{(t+1)}-\xx^{(t+1)}}_{\infty}&\\
    &\le \frac{4\eps(1-\alpha_{\min})d^{1/2}_{\max}}{d^{1/2}_{\min}}\sqrt{n} + \kh{1-\alpha_{\min}}\cdot\norm{\xxtil^{(t)}-\xx^{(t)}}_{\infty}\\
    &\le \frac{4\eps\sqrt{nd_{\max}d^{-1}_{\min}}\cdot\kh{1-\alpha_{\min}}\left[1-\kh{1-\alpha_{\min}}^{t+1}\right]}{\alpha_{\min}},
\end{align*}
as required.
\end{proof}

In order to show the convergence of this method, it needs to prove that, after sufficiently many iterations, the error between $\xx^{(t)}$ and $\zz^\ast$ will be sufficiently  small, as characterized by the following lemma.
\begin{lemma}
\label{lem:xt}
For every $t\ge0$,
\begin{equation}
    \norm{\xx^{(t)}-\zz^\ast}_{\infty} \le \frac{(1-\alpha_{\min})^t}{\alpha_{\min}}.
\end{equation}
\end{lemma}
\begin{proof}
Expanding with the series $\kh{\II-\kh{\II-\ALPHA}\PP^\ast}^{-1}=\sum\limits_{j=0}^{\infty} \kh{\kh{\II-\ALPHA}\PP^\ast}^j$ leads to
\begin{equation}
    \zz^\ast-\xx^{(t)}=\sum_{j=t}^{\infty}\kh{\kh{\II-\ALPHA}\PP^\ast}^j\ALPHA\ss - \kh{\kh{\II-\ALPHA}\PP^\ast}^t\ss.
\end{equation}

Below, by induction, it will be shown that for any $\xx\in[0,1]^n$, the relation $\smallnorm{\kh{\kh{\II-\ALPHA}\PP^\ast}^j\xx}_{\infty}\le(1-\alpha_{\min})^j$ holds for all $j\ge 0$.
Since every coordinate of $\xx$ lies in the interval $[0,1]$, it is obvious that the above relation is true for the case of $j=0$.
Suppose that, for some $j>0$, every coordinate of $\yy=\kh{\kh{\II-\ALPHA}\PP^\ast}^j\xx$ has magnitude at most $(1-\alpha_{\min})^j$.
Since $\PP^\ast$ is row-stochastic, it follows that $\norm{\PP^\ast\yy}_{\infty}\le(1-\alpha_{\min})^j$.
In addition, because $\alpha_i\ge\alpha_{\min}$ for all $i\in\calV$, one has $\norm{\kh{\II-\ALPHA}\PP^\ast\yy}_{\infty}\le(1-\alpha_{\min})^{j+1}$, completing the induction proof.

Finally, since both $\sum_{j=t}^{\infty}\kh{\kh{\II-\ALPHA}\PP^\ast}^j\ALPHA\ss$ and $\kh{\kh{\II-\ALPHA}\PP^\ast}^t\ss$ have non-negative coordinates, one has
\begin{align*}
    \norm{\xx^{(t)}-\zz^\ast}_{\infty}
    &\le \max\Bigg\{\norm{\sum_{j=t}^{\infty}\kh{\kh{\II-\ALPHA}\PP^\ast}^j\ALPHA\ss}_{\infty},
	\norm{\kh{\kh{\II-\ALPHA}\PP^\ast}^t\ss}_{\infty}\Bigg\} \\
    \le \sum_{j=t}^{\infty} (1-\alpha_{\min})^j
    &=\frac{(1-\alpha_{\min})^t}{\alpha_{\min}},
\end{align*}
as claimed by the lemma.
\end{proof}

Combining Lemmas~\ref{lem:xtil} and~\ref{lem:xt}, a convergent approximate iteration method can be summarized as stated in the following theorem.
\begin{theorem}[Approximation Error]
	\label{th:ae}
For every $t\ge0$,
\begin{align*}
    \norm{\xxtil^{(t)}-\zz^\ast}_{\infty}
    \le
\frac{4\eps\sqrt{n}\cdot\kh{1-\alpha_{\min}}\left[1-\kh{1-\alpha_{\min}}^t\right] + (1-\alpha_{\min})^t}{\alpha_{\min}}.
\end{align*}\label{ThAppr}
\end{theorem}

In the sequel, this approximate iteration algorithm is referred to as \textsc{Approx}. It should be mentioned that Theorem~\ref{ThAppr} provides only a rough upper bound. The experiments in Section~\ref{sec:result} show that \textsc{Approx} works well in practice, leading to very accurate results for real networks.

\section{Experiments on Real Networks}
\label{sec:exp}

In this section, we conduct extensive experiments on real-world social networks to evaluate the performance of the algorithm \textsc{Approx}. % We introduce our experimental setup and datasets in Section~\ref{sec:setup}. We compare the behavior of second-order FJ model with that of the standard FJ model on several real-world networks in Section~\ref{sec:comp}. Finally, we report the results for performance evaluation and parameter analyses in Section~\ref{sec:result} and Section~\ref{sec:pa}, respectively.

%%%%%%%%%%%%%%%%%%%%%%%%%%%%%%%%%%%%%%%%%%%%%%%%%%%
\begin{table}
	\centering
	\caption{Statistics of real networks used in experiments and comparison of running time (seconds, $s$) between \textsc{Exact} and \textsc{Approx} for three innate opinion distributions (uniform distribution, exponential distribution, and power-law distribution). }\label{T1}
	\resizebox{\linewidth}{!}{
		\begin{tabular}{ccccclcclcc}
			\toprule
			%\multirow{2}*{Network} & \multirow{2}*{$n'$} & \multirow{2}*{$m'$} &
			&&&&\multicolumn{6}{c}{Running time ($s$) for \textsc{Exact} and \textsc{Approx} algorithms } \\
			\cmidrule{4-11}
			\multirow{2}*{Network} & \multirow{2}*{$n'$} & \multirow{2}*{$m'$} &\multicolumn{2}{c}{Uniform distribution}
			&& \multicolumn{2}{c}{Exponential distribution}
			&& \multicolumn{2}{c}{Power-law distribution} \\
			\cmidrule{4-5}
			\cmidrule{7-8}
			\cmidrule{10-11}
			& & &   \textsc{Exact}& \textsc{Approx}&& \textsc{Exact} & \textsc{Approx}&& \textsc{Exact}& \textsc{Approx} \\
			\midrule
			%GrQc & 4158 & 13422  & 2.049 & 2.351 &2.151 &0.116 & 2.369& 0.115\\
			HamstersterFriends & 1788 & 12476  & 0.174 &0.974&&0.158&0.876 &&0.176 & 0.866\\
			HamstersterFull & 2000 & 16098  &  0.303& 1.540 &&0.316 &1.568 &&0.317 & 1.547 \\
			PagesTVshow & 3892 & 17239  &  1.204& 1.530 &&1.126 &1.367 &&1.083 & 1.367 \\
			Facebook (NIPS) & 4039 & 88234  &  1.492& 6.274 &&1.473 &6.331 &&1.556 & 6.243 \\
			PagesGovernment & 7057 & 89429  &5.857  &7.679  && 5.682 & 7.316&& 5.682 &7.353 \\
			Anybeat & 12645 & 49132  & 31.448 &  4.730&& 31.843& 5.462&& 31.575& 4.739 \\
			PagesCompany & 14113 & 52126  & 39.348 & 4.269 && 37.690& 3.905&& 37.477& 3.877\\
			Gplus & 23613 & 39182  & 163.525 &4.329  &&171.166 & 4.295&& 165.470& 4.307\\
			GemsecRO & 41773 & 125826  &885.069  & 15.758 && 888.009& 15.519&& 873.696&16.079  \\
			GemsecHU & 47538 & 222887  & 946.399 & 28.592 &&937.540 & 30.562&& 919.438&29.174  \\
			PagesArtist & 50515 & 819090  & 1160.469 & 139.565 &&1183.787 & 138.507&& 1167.293& 148.941 \\
			Brightkite & 56739 & 212945  & 1913.246 & 27.351 && 1901.307& 29.203&&1895.432 & 28.361 \\
			Livemocha* & 104103 & 2193083  & --- & 538.730 && ---& 542.467&& ---&542.708  \\
			Douban* & 154908 & 327162  & --- & 44.166 && ---& 43.292&&--- & 43.350 \\
			Gowalla* & 196591 & 950327  & --- &  138.222&& ---&142.690 && ---&143.669  \\
			TwitterFollows* & 404719 & 713319  &---  &  96.850&& ---& 97.319&& ---&96.748  \\
			Delicious* & 536108 & 1365961  & --- & 209.371 &&--- &206.960 && ---& 206.739 \\
			YoutubeSnap* & 1134890 & 2987624  & --- & 663.090 && --- & 667.921&&--- & 667.770 \\
			Hyves* & 1402673 & 2777419  & --- & 648.906 && ---& 633.172&& ---&636.219  \\
			\bottomrule
		\end{tabular}
	}
\end{table}
%%%%%%%%%%%%%%%%%%%%%%%%%%%%%%%%%%%%%%%%%%%%%%%%%%%

\subsection{Setup}
\label{sec:setup}

\textbf{Machine Configuration and  Reproducibility.}
Our extensive experiments run on a Linux box with 16-core 3.00GHz Intel Xeon E5-2690 CPU and 64GB of main memory.
All algorithms are programmed in \emph{Julia v1.3.1}.
The source code is publicly available at \url{https://github.com/HODynamic/HODynamic}.

\noindent\textbf{Datasets.}
We test the algorithm on a large set of realistic networks, all of which are collected from the Koblenz Network Collection~\citep{Ku13} and Network Repository~\citep{RoAh15}. For those networks that are disconnected originally, we perform experiments on their largest connected components. The statistics of these networks are summarized in the first three columns of Table~\ref{T1}, where we use $n'$ and $m'$ to denote, respectively, the numbers of nodes and edges in their largest connected components. The smallest network consists of $4,991$ nodes, while the largest network has more than one million nodes. In Table~\ref{T1}, the networks are listed in an increasing order of the number of nodes in their largest connected components.

\noindent\textbf{Input Generation.}
For each dataset, we use the network structure to generate the input parameters in the following way. The innate opinions are generated according to three different distributions, that is, uniform distribution, exponential distribution, and power-law distribution, where the latter two are generated by the randht.py file in~\citep{ClShNe09}. For the uniform distribution, we generated the opinion $s_i$ of node $i$ at random in the range of $[0,1]$. For the exponential distribution, we use the probability density $\mathrm{e}^{ x_{\min }}\mathrm{e}^{-x}$ to generate $n'$ positive real numbers $x$ with minimum value $x_{\min}>0$. Then, we normalize these $n'$ numbers to be within the range $[0,1]$ by dividing each $x$ with the maximum observed value. Similarly, for the power-law distribution, we choose the  probability density $(\alpha-1)x_{\min}^{\alpha-1}x^{-\alpha}$ with $\alpha=2.5$ to generate $n'$ positive real numbers, and then normalize them to be within the interval $[0,1]$ as the innate opinions. 
In practice, we set $x_{\min}=1$ for both the exponential and power-law distribution.
We note that there is always a node with innate opinion $1$ due to the normalization operation for the latter two distributions. We generate the resistance parameters uniformly to be within the interval $(0,1)$.

%\subsection{Comparison of Equlibrium Opinions between Standard FJ Model and Second-Order FJ Model}

\subsection{Comparison between Standard FJ Model and Second-Order FJ Model}
\label{sec:comp}

To show the impact of higher-order interactions on the opinion dynamics, we compare the equilibrium expressed opinions between the second-order FJ model and the standard FJ model on four real networks: PagesTVshow, PagesCompany, Gplus, and GemsecRO. For both models, we generate innate opinions and resistance parameters for each node according to the uniform distribution. We set $\beta_1=1, \beta_2=0$ for the standard FJ model, and $\beta_1=0, \beta_2=1$ for the second-order FJ model.

%%%%%%%%%%%%%%%%%%%%%%%%%%%%%%%%%%%%%%%%%%%%%%%%%%%
\begin{figure}[ht]
	\centering
	\includegraphics[width=.6\linewidth]{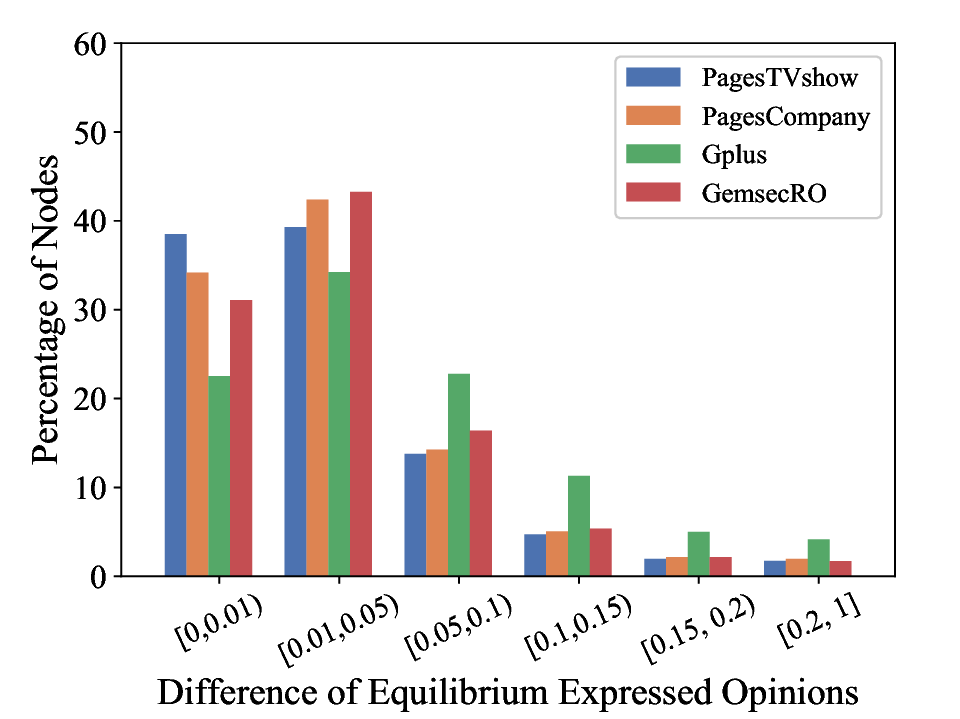}
	\caption{Distribution for difference of equilibrium expressed opinions between the standard FJ model and the second-order FJ model on four real networks.	}	\label{fig:deviation}
\end{figure}
%%%%%%%%%%%%%%%%%%%%%%%%%%%%%%%%%%%%%%%%%%%%%%%%%%%

Figure~\ref{fig:deviation} illustrates the distribution for the difference of the final expressed opinions for each node between the classic and second-order FJ models on four considered real networks. It can be observed that for each of these four  networks, there are more than half nodes, for which the difference of expressed opinions between the two models is larger than $0.01$. Particularly, there are over $10\%$ agents, for which the difference of equilibrium opinions is greater than $0.1$. This possibly makes them stand on the opposite sides for different models. Thus, the opinion dynamics for the second-order FJ model differs largely from the classic FJ model, indicating that the effects of higher-order interactions are not negligible.

\subsection{Performance Evaluation}
\label{sec:result}

To evaluate the performance of the new algorithm \textsc{Approx}, we implement it on various real networks and compare the running time and accuracy of \textsc{Approx} with those corresponding to the standard \textsc{Exact} algorithm. For the \textsc{Exact}, it computes the equilibrium vector by calculating the random-walk matrix polynomials via matrix multiplication and directly inverting the matrix $\II-(\II-\ALPHA)\PP^\ast$. Here, we use the second-order random-walk matrix polynomial to simulate the opinion dynamics with $\beta_1=\beta_2=0.5$. For \textsc{Approx}, we set the number $M$ of samples as $10\times T\times m$ and approximate the equilibrium vector with $100$ iterations. 
By using the maximum iterations as the stopping criterion, we can avoid the numeric errors and instability caused by other stopping criteria and thus make the efficiency comparison more stable.
To objectively evaluate the running time, we enforce the program to run on a single thread for both \textsc{Exact} and \textsc{Approx} on all considered networks, except the last seven marked with asterisks, for which we cannot run \textsc{Exact} due to the very high cost for space and time.

\noindent\textbf{Efficiency.}
We present the running time of  algorithms \textsc{Approx} and \textsc{Exact} for all networks in Table~\ref{T1}. For the last seven networks, we only run algorithm \textsc{Approx} since \textsc{Exact} would take extremely long time. For each of the three innate opinion distributions in different networks, we record the running time of \textsc{Approx} and \textsc{Exact}. From Table~\ref{T1}, we observe that for small networks with less than 10,000 nodes, the running time of \textsc{Approx} is a little longer than that of  \textsc{Exact}. Thus, \textsc{Approx} shows no superiority for small networks. However, for those networks having more than twenty thousand nodes, \textsc{Approx} significantly improves the computation efficiency compared with \textsc{Exact}. For example, for the moderately large network GemsecRO with 41,773 nodes, \textsc{Approx} is $60\times$ faster than \textsc{Exact}. Finally, for large graphs \textsc{Approx} shows a very obvious efficiency advantage. Table~\ref{T1} indicates that for those networks with over 100 thousand nodes, \textsc{Approx} completes running within $12$ minutes, whereas \textsc{Exact} fails to run. We note that for large networks, the running time of \textsc{Approx} grows nearly linearly with respect to $m'$, consistent with the above complexity analysis, while the running time of \textsc{Exact} grows as a cube power of $n'$.

\noindent\textbf{Accuracy.}
In addition to the high efficiency, the new algorithm \textsc{Approx} provides a good approximation for the equilibrium opinion $ \zz^* =( \zz_1^*,\zz_2^*, ..., \zz_{n}^*)^\top$ in practice. To show this, we compare the approximate results of \textsc{Approx} for second-order FJ model with exact results obtained by \textsc{Exact}, for all the examined networks shown in Table~\ref{T1}, except the last seven which are too big for \textsc{Exact} to handle. For each of the three distributions of the innate opinions, Table~\ref{T2} reports the mean absolute error $\sigma = \sum\nolimits_{i=1}^{n'} |\zz^*_i - \zztil^*_i|/n'$, where $\zztil^*=( \zztil_1^*,\zztil_2^*,..., \zztil_{n}^*)^\top$ is the estimated vector obtained by \textsc{Approx}. From Table~\ref{T2}, we observe that the actual mean absolute errors $\sigma$ are all less than $0.008$, thus ignorable. Furthermore, for all networks we tested, the mean absolute errors $\sigma$ are smaller than the theoretical ones provided by Theorem~\ref{th:ae}. Therefore, the new algorithm \textsc{Approx} provides a very desirable approximation for the equilibrium opinion vector in applications.

%%%%%%%%%%%%%%%%%%%%%%%%%%%%%%%%%%%%%%%%%%%%%%%%%%%
\begin{table}[htbp]
	\centering
	\begin{threeparttable}
	\small
	\caption{Mean absolute error ($\times 10^{-3}$) for estimated expressed opinions of the second-order FJ model with respect to three innate opinion distributions.% (uniform distribution, exponential distribution, and power-law distribution).
	}\label{T2}
	%\resizebox{\linewidth}{!}{
		\begin{tabularx}{0.7\linewidth}{cccc}
			\toprule
			% && \multicolumn{5}{c}{Mean absolute error} \\
			%\cmidrule{3-7}
			Network & Uniform & Exponential & Power-law \\
			%\cmidrule{2-3}
			%\cmidrule{5-6}
			%&  \textsc{Iter} & \textsc{Solve} & & \textsc{Iter} & \textsc{Solve} \\
			\midrule
			HamstersterFriends    & 3.718 & 1.738  &   0.123   \\
			HamstersterFull    & 3.134 & 1.315  &   0.311   \\
			PagesTVshow  &   4.552 & 1.553  & 0.226   \\
			Facebook (NIPS)    & 1.967 & 0.689  &   0.125   \\
			PagesGovernment    & 2.622 & 0.885  & 0.047   \\
			Anybeat  &   5.896 & 2.504   & 0.019   \\
			PagesCompany    & 4.788 & 1.575   & 0.137   \\
			Gplus &   7.152 & 1.638  &   0.071   \\
			GemsecRO   & 4.787 & 1.165  &   0.048   \\
			GemsecHU   & 3.611 & 0.960  &   0.031   \\
			PagesArtist  & 2.746 & 0.669  &  0.113  \\		
			Brightkite  & 5.717 & 1.756  &  0.057  \\		
			
			\bottomrule
		\end{tabularx}
	%}	
	\end{threeparttable}
\end{table}
%%%%%%%%%%%%%%%%%%%%%%%%%%%%%%%%%%%%%%%%%%%%%%%%%%%

\subsection{Parameter Analysis}
\label{sec:pa}

We finally discuss how the parameters affect the performance and efficiency of \textsc{Approx}. We report all the parameter analyses on four networks, namely HamstersterFriends, HamstersterFull, PagesTVshow, and Facebook (NIPS). All experiments here are performed for the second-order FJ opinion  dynamics model with $\beta_1=\beta_2=0.5$.

%%%%%%%%%%%%%%%%%%%%%%%%%%%%%%%%%%%%%%%%%%%%%%%%%%%
\begin{figure}[ht]
	\centering
	\includegraphics[width=0.6\linewidth]{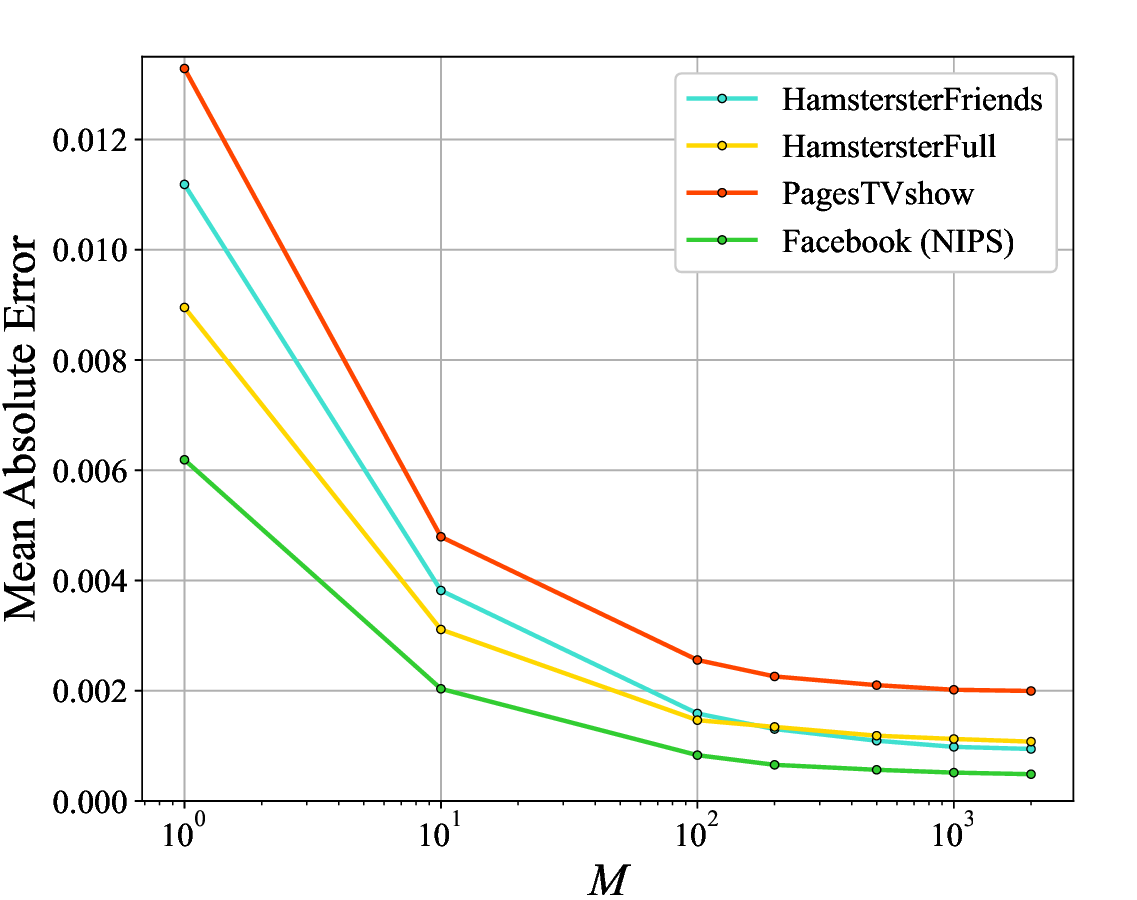}
	\caption{Mean absolute error} v.s. the number of non-zeros $M$. 	\label{fig:M}
\end{figure}
%%%%%%%%%%%%%%%%%%%%%%%%%%%%%%%%%%%%%%%%%%%%%%%%%%%

\noindent\textbf{The number of non-zeros $M$.}
As shown in Section~\ref{sec:sgs}, $M = O(Tm\eps^{-2}\log n)$ is required to guarantee the approximation error in theory. We first explore how the number of non-zeros influences the performance of the algorithm \textsc{Approx} in implementation. Without loss of generality, we empirically set $M$ to be $k\times T\times m$, with $k$ being $1, 10, 100, 200, 500, 1000$ and $2000$, respectively. In Figure~\ref{fig:M}, we report the mean absolute error of \textsc{Approx}, which drops as we increase the number of samples $M$. This is because matrix $\PP^\ast$ is approximated more accurately for larger $M$. On the other hand, although increasing $M$ may have a positive influence on the accuracy of \textsc{Approx}, this marginal benefit diminishes gradually. Figure~\ref{fig:M} shows that  $M = 10\times T\times m$  is in fact a desirable choice, which balances the trade-off between the effectiveness and efficiency of \textsc{Approx}.

%%%%%%%%%%%%%%%%%%%%%%%%%%%%%%%%%%%%%%%%%%%%%%%%%%%
\begin{figure}[ht]
	\centering
	\includegraphics[width=0.6\linewidth]{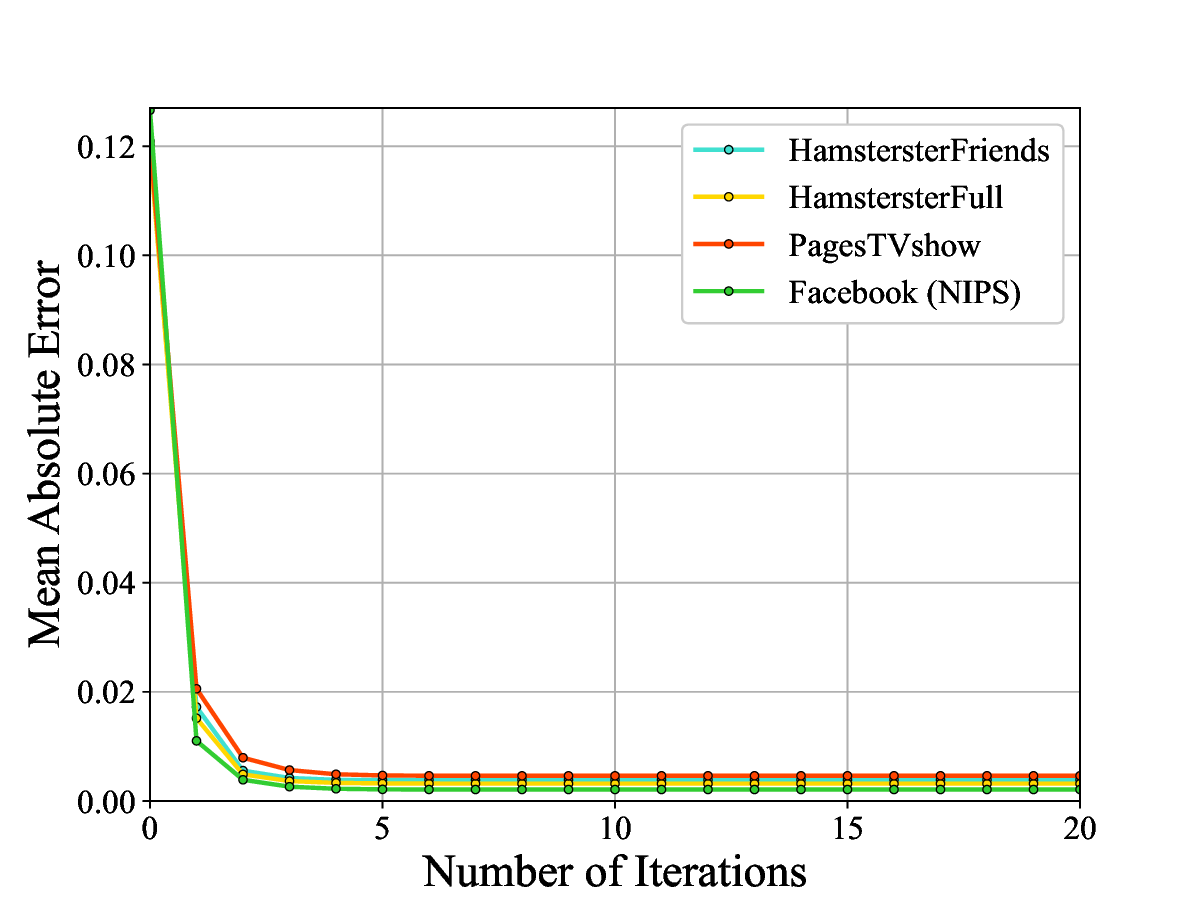}
	\caption{Mean absolute error v.s. the number of iterations. 	\label{fig:Iters}}
\end{figure}
%%%%%%%%%%%%%%%%%%%%%%%%%%%%%%%%%%%%%%%%%%%%%%%%%%%

\noindent\textbf{The number of iterations on small networks.}
In Section~\ref{sec:iter}, we present an approximation convergence for the iteration method \textsc{Approx}, with the accuracy of \textsc{Approx} depending on the number of iterations. In Figure~\ref{fig:Iters}, we plot the mean absolute error of \textsc{Approx} as a function of the number of iterations. In all experiments, $M$ is set to be $10\times T\times m$. As demonstrated in Figure~\ref{fig:Iters}, the second-order FJ model converges in several iterations for all the four networks tested. 

%%%%%%%%%%%%%%%%%%%%%%%%%%%%%%%%%%%%%%%%%%%%%%%%%%%
\begin{figure}[ht]
	\centering
	\includegraphics[width=0.6\linewidth]{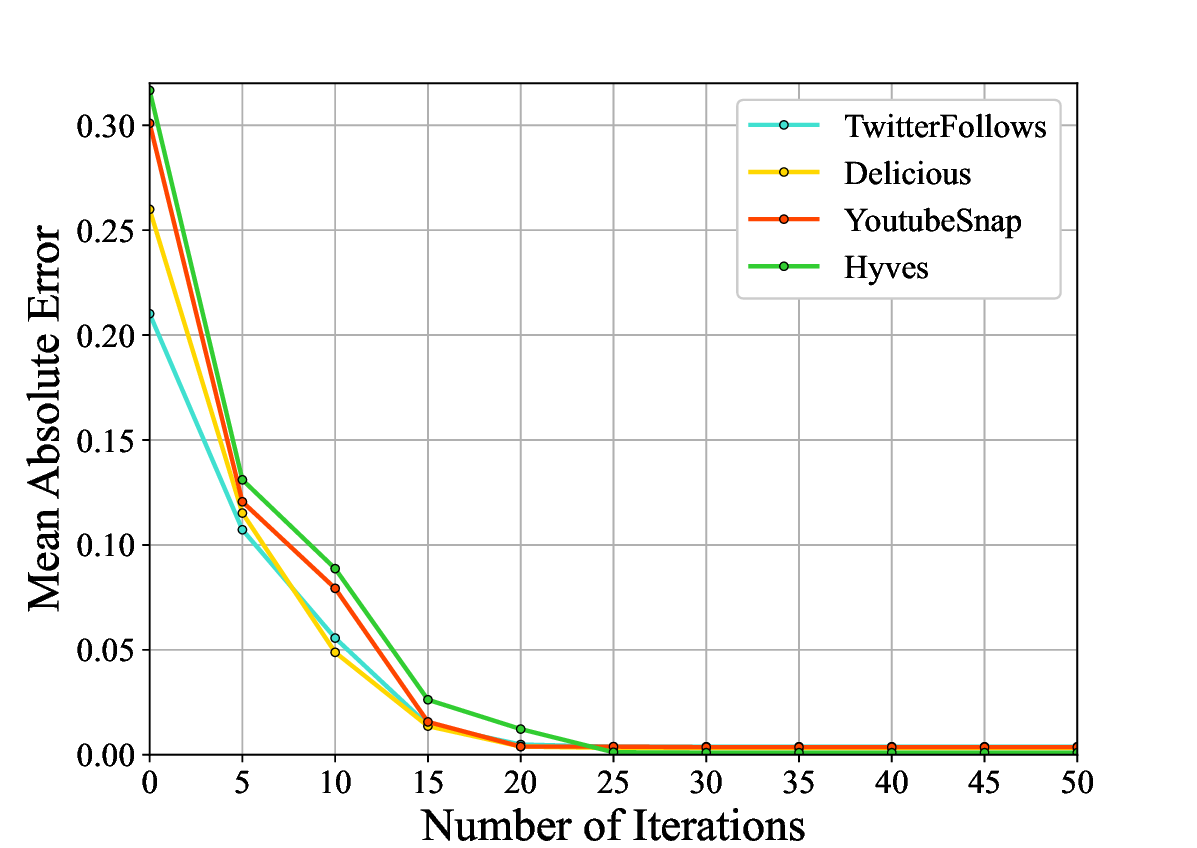}
	\caption{Mean absolute error between two iterations v.s. the number of iterations. 	\label{fig:ItersL}}
\end{figure}
%%%%%%%%%%%%%%%%%%%%%%%%%%%%%%%%%%%%%%%%%%%%%%%%%%%

\noindent\textbf{The number of iterations on large networks.}
We also analyze the influence of the number of iterations on large networks. 
Since we cannot run \textsc{Exact} for these four large networks, we instead use the mean absolute error between two iterations in \textsc{Approx} as an indicator for convergence in Figure~\ref{fig:ItersL}.
In all experiments, $M$ is set to be $10\times T\times m$. 
As demonstrated in Figure~\ref{fig:ItersL}, the second-order FJ model converges in dozens of iterations for all the four networks tested. 
Therefore, one hundred iterations are enough to obtain desirable approximation results for networks with millions of nodes. 
For all the experiments shown in Table~\ref{T1}, the difference between opinion vectors of two consecutive iterations for the second-order FJ model is insignificant after dozens of iterations.

\section{Conclusion}

In this paper, we presented a significant extension of the classic Friedkin-Johnsen (FJ) model by considering not only nearest-neighbor interactions, but also long-range interactions via leveraging higher-order random walks. We showed that the proposed model has a unique equilibrium expressed opinion vector, provided that each individual holds an innate opinion. %, which has some influence on the expressed opinions.
We also demonstrated that the resultant expressed opinion vector of the new model may be significantly different from that of the FJ model, indicating the important impact of higher-order interactions on opinion dynamics.

The expressed opinion vector of the new model can be considered as an expressed opinion vector of the FJ model in a dense graph with a loop at every node, whose transition matrix is a convex combination of  powers of the transition matrix for the original graph. However, direct computation of the transition matrix for the dense graph is computationally expensive, which involves multiple matrix multiplication and inversion operations. As a remedy, {we leveraged the state-of-the-art Laplacian sparsification technique and  the nearly linear-time algorithm in~\cite{ChChLiPeTe15b} 
to obtain a sparse matrix, which is spectrally similar to the original dense matrix thereby preserving all basic information.}  Based on the obtained sparse matrix, we further proposed a convergent iteration algorithm, which approximates the equilibrium opinion vector in linear space and time. We finally conducted extensive experiments on diverse social networks, which demonstrate that the new algorithm  achieves both good efficiency and effectiveness. %, which can solve the expressed opinions of high-order model on large graphs containing millions of nodes.

{It should be mentioned that in this paper, we only focus on the impacts of higher-order interactions on the sum of expressed opinions. Actually, in addition to the opinion sum, there are many other related quantities for opinion dynamics, including convergence rate, polarization~\cite{DaGoLe13,MaTeTs17, MuMuTs18}, disagreement~\cite{MuMuTs18,ZhBaZh21}, and so on. It is expected that higher-order interactions have also important influences on these important quantities. Although these subjects are beyond our paper, below we provide a heuristic explanation for the reason of  higher-order interactions affecting polarization. As shown in Theorem~\ref{theorem:Zstar}, the vector of expressed opinions is determined simultaneously by three factors: the innate opinion $s_i$ and resistance parameter $\alpha_i$ of every agent $i$, as well as the higher-order interaction encoded in matrix $P^*$. Thus, higher-order interactions play a significant role in opinion polarization, since this quantity is also simultaneously affected by these three factors~\cite{MuMuTs18}. Finally, it is worth emphasizing that although our model incorporates higher-order interactions and thus generates an opinion vector different from that of the classic FJ model, it is difficult to judge which opinion vector is superior or more compelling. In fact, these two models are not mutually exclusive. The choice of the models depends on the specific aim of applications, such as minimizing polarization~\cite{MuMuTs18}, disagreement~\cite{ZhBaZh21}, or conflict~\cite{ZhZh22,WaKl23}.}

%In future work, we will consider how this kind of higher-order interactions affecting  the social phenomena, such as opinion polarization and controversy~\cite{ZhBaZh21}.

\section*{Acknowledgment}

This work was supported by the National
	Natural Science Foundation of China (Nos. 62372112, U20B2051, and 61872093).

%This work was supported in part by the National Natural Science Foundation of China (Nos. U20B2051, and 61872093), the National Key R \& D Program of China (Nos. 2018YFB1305104 and 2019YFB2101703),  the Shanghai Municipal Science and Technology Major Project (No.  2018SHZDZX01), and ZJLab.  and the Innovation Action Plan of Shanghai Science and Technology (Nos. 20222420800 and 20511102200). 

\bibliography{Opinion}   % name your BibTeX data base

\end{document}